\documentclass[twocolumn,aps,superscriptaddress,showpacs]{revtex4}
\usepackage{amssymb}
\usepackage{amsmath,bm}
\usepackage{graphicx}
\usepackage[normalem]{ulem}
\usepackage[dvips]{color}
\usepackage{mhchem} 
\renewcommand\sout{\bgroup \color{red} \ULdepth=-.5ex \ULset}

\hfuzz=\maxdimen
\begin{document}

\title{Low density nuclear matter with light clusters in a generalized nonlinear relativistic mean-field model}
\author{Zhao-Wen Zhang}
\affiliation{School of Physics and Astronomy and Shanghai Key Laboratory for
Particle Physics and Cosmology, Shanghai Jiao Tong University, Shanghai 200240, China}
\author{Lie-Wen Chen\footnote{%
Corresponding author (email: lwchen$@$sjtu.edu.cn)}}
\affiliation{School of Physics and Astronomy and Shanghai Key Laboratory for
Particle Physics and Cosmology, Shanghai Jiao Tong University, Shanghai 200240, China}
\affiliation{Center of Theoretical Nuclear Physics, National Laboratory of Heavy Ion
Accelerator, Lanzhou 730000, China}
\date{\today}

\begin{abstract}
We systematically investigate the thermodynamic properties of homogeneous nuclear matter with light clusters at low densities and finite temperatures using a generalized nonlinear relativistic mean-field (gNL-RMF) model, in which light clusters up to $\alpha $ ($1 \le A \le 4$) are included as explicit degrees of freedom and treated as point-like particles with their interactions described by meson exchanges and the medium effects on the cluster binding energies are described by density- and temperature-dependent energy shifts with the parameters obtained by fitting the experimental cluster Mott densities. We find that the composition of low density nuclear matter with light clusters is essentially determined by the density- and temperature-dependence of the cluster binding energy shifts. Compared with the values of the conventional (second-order) symmetry energy, symmetry free energy and symmetry entropy, their fourth-order values are found to be significant at low densities ($n \sim 10^{-3}$ fm$^{-3}$) and low temperatures ($T \lesssim 3$ MeV), indicating the invalidity of the empirical parabolic law for the isospin asymmetry dependence of these nuclear matter properties. Our results indicate that in the density region of $n \gtrsim  0.02$ fm$^{-3}$, the clustering effects become insignificant and the nuclear matter is dominated by nucleon degree of freedom. In addition, we compare the gNL-RMF model predictions with the corresponding experimental data on the symmetry energy and symmetry free energy at low densities and finite temperatures extracted from heavy-ion collisions, and reasonable agreement is found.
\end{abstract}

\maketitle
\section{Introduction}
Understanding the properties of nuclear matter, especially its equation of state (EOS), is extremely important in nuclear physics and astrophysics. In order to investigate various nuclear and astrophysical phenomena such as heavy-ion collisions, supernova explosion, and neutron star evolution, we need the exact information on nuclear matter EOS in a wide range of temperature, density, and isospin asymmetry~\cite{Dan02,Lat04}. In the last few decades, significant progress has been made in determining the nuclear matter EOS from experiments, observations and theoretical calculations~\cite{Dan02,Lat04,Ste05,Bar05,LCK08,Heb15,Car15,Oer17}. However, knowledge on the EOS of isospin asymmetric nuclear matter, especially the nuclear symmetry energy $E_{\rm sym}(n)$ at densities far from saturation density $n_0$ of symmetric nuclear matter, is still poorly known (see, e.g., Ref.~\cite{LiBA14} for a recent review).

In nuclear matter at very low densities, it is known that nucleons tend to form light nuclei to reduce energy of the system~\cite{Rop09,Typ10,Typ13,Hag14,Giu14}. Such a clustering phenomenon may exist in the crust of neutron stars~\cite{Fat10,Bha10,Ava12,Rad14,Ava17}, in core-collapse supernovae~\cite{Fis14,Fur17}, and even in heavy-ion collisions~\cite{ChenLW03,Gai04,ZhangYX12,Rop13,Hem15}. It is thus interesting to consider the clustering effects on the properties of nuclear matter under various conditions of density, temperature and isospin asymmetry.
Recently, great efforts have been made for investigating the clustering effects. For example, the clustering effects on nuclear structure properties have been explored and some novel features are revealed~\cite{Ebr12,Zho13,Yan14,He14,Aym14,THSR17}.
In addition, in some commonly used EOSs for the simulation of core-collapse supernova, e.g., the Lattimer-Swesty EOS constructed by Lattimer and Swesty~\cite{Lat91} and the Shen EOS constructed by Shen \textit{et al}.~\cite{She11}, the $\alpha$-particles are included and treated as an ideal Boltzmann gas.
The EOS of low density nuclear matter including nucleons and $\alpha$-particles is also investigated by using the virial expansion~\cite{Hor06}, and later on the contributions of deuteron ($d=^{2}$H), triton ($t=^{3}$H) and helium-3 ($h=^{3}$He) as well as heavier nuclei are further included and investigated by using the S-matrix method and the quasiparticle gas model~\cite{Mal08,Hec09}.

During the last several years, Typel \textit{et al.} have investigated nuclear matter including formation of light clusters up to the $\alpha$-particle by using a generalized density-dependent relativistic mean-field (gDD-RMF) model~\cite{Typ10}, in which the density- and temperature-dependence of the binding energy of clusters in nuclear medium is obtained from the predictions of quantum statistical (QS) approach. The density-dependence of the cluster binding energies and coupling parameters brings the ``rearrangement'' contributions in the particle equations of motion and vector self-energies~\cite{Had93,Len95,Ces98,Typ99,Roc11}.
Since the work of Typel \textit{et al.}~\cite{Typ10}, a number of studies have been carried out to explore the clustering effects in nuclear matter. For instance, Sharma \textit{et al}.~\cite{Bha10} study the clustering effects on the liquid-gas phase transition, composition, and structure of protoneutron stars including hyperons. Avancini \textit{et al.}~\cite{Ava12} investigate the property of nuclear pasta phase including the $\alpha$-particles and other light nuclei. Ferreira \textit{et al.}~\cite{Fer12} fit the density-dependence of the in-medium cluster binding energies of Ref.~\cite{Typ10} by changing the couplings of light clusters in the RMF model, and explore the properties of light clusters in nuclear matter. For simplicity, the binding energies of light clusters in nuclear medium are usually treated as constants in these works~\cite{Ava12,Fer12}.

Experimentally, clustering effects in low density nuclear matter have been investigated in heavy-ion collisions around Fermi energies. In particular, information of nuclear matter symmetry energy at low densities and finite temperatures have been extracted in heavy-ion collision experiments by Natowitz \textit{et al.}~\cite{Nat10}, Kowalski \textit{et al.}~\cite{Kow07}, and Wada \textit{et al.}~\cite{Wad12}. The results indicate that the clustering effects can enhance drastically the symmetry energy at low densities, while the conventional mean-field models without considering clusters significantly under-predict the experimentally measured values of the symmetry energy. Meanwhile, the experimental data~\cite{Hag12} on the dissolution density (Mott density) of clusters in nuclear medium are also obtained.

In the present work, we investigate the properties of nuclear matter with light clusters at low densities and finite temperatures by using a generalized nonlinear relativistic mean-field (gNL-RMF) model. We also discuss the Mott density of clusters in nuclear matter as well as compare the model predictions with the experimental data on the symmetry energy and symmetry free energy extracted from heavy-ion collisions.
In the gNL-RMF model, light clusters up to $\alpha $ ($1 \le A \le 4$) are included as explicit degrees of freedom and treated as point-like particles with their interactions described by meson exchanges, and the medium effects on the light cluster binding energies are described by density- and temperature-dependent energy shifts.
In the non-linear RMF (NL-RMF) model~\cite{Lal97,Hor01,Tod03,Tod05,Pie06,Cai12}, the nonlinear couplings of mesons are introduced to reproduce the ground-state properties of finite nuclei and to modify the density dependence of the symmetry energy $E_{\rm sym}(n)$. Since all of these couplings are constants in the NL-RMF model, the calculations are thus more convenient and simpler.

This paper is organized as follows. In Section~\ref{Model}, we introduce the gNL-RMF model for low density nuclear matter including light clusters. And then the theoretical results are presented and some experimental data are compared with the theoretical predictions in Section~\ref{Result}. Finally we give a conclusion in Section~\ref{Summary}.

\section{Theoretical framework}
\label{Model}

We extend the NL-RMF model and study the properties of a homogeneous nuclear matter system with multi-components including protons, neutrons and light clusters of $d$, $t$, $h$ and $\alpha$.
All the components are treated as point-like particles, and they interact through the exchange of various effective mesons including isoscalar scalar ($\sigma$) and vector ($\omega$) mesons and an isovector vector ($\rho$) meson.
In this gNL-RMF model, the Lagrangian density of the system reads
\begin{eqnarray}
\mathcal{L} &=& \sum_{i=p, n, t, h} \mathcal{L}_i + \mathcal{L}_{\alpha} + \mathcal{L}_d
+ \mathcal{L}_{\rm{meson}},
\label{eq:LRMF}
\end{eqnarray}%
where the fermions ($i=p, n, t, h$) with spin $1/2$ are described by
\begin{eqnarray}
\mathcal{L}_i = \bar{\Psi}_i\left[\gamma_{\mu}iD^{\mu}_i-M^{*}_i\right]\Psi_i,
\label{eq:Lj}
\end{eqnarray}%
while the Lagrangian densities of $\alpha$-particle with spin 0 and deuteron with spin $1$ are given, respectively, by
\begin{eqnarray}
\mathcal{L}_{\alpha} &=& \frac{1}{2}\left(iD^{\mu}_{\alpha}\phi_{\alpha}\right)^{*}\left(iD_{\mu\alpha}\phi_{\alpha}\right)
-\frac{1}{2}\phi^{*}_{\alpha}\left(M^{*}_{\alpha}\right)^2\phi_{\alpha},
\label{eq:La}
\end{eqnarray}%
and
\begin{eqnarray}
\mathcal{L}_d &=& \frac{1}{4}\left(iD^{\mu}_d\phi^{\nu}_d-iD^{\nu}_d\phi^{\mu}_d\right)^{*}
\left(iD_{d\mu}\phi_{d\mu}-iD_{d\nu}\phi_{d\nu}\right) \nonumber\\
&-&\frac{1}{2}\phi^{\mu*}_d\left(M^{*}_d\right)^2\phi_{d\mu}.
\label{eq:Ld}
\end{eqnarray}%
The covariant derivative is defined by
\begin{eqnarray}
iD^{\mu}_i=i\partial^{\mu}-A_i g_{\omega}\omega^{\mu}
-\frac{g_{\rho}}{2}\overrightarrow{\tau}\cdot\overrightarrow{\rho}^{\mu},
\label{eq:iDj}
\end{eqnarray}%
and the effective mass is expressed as
\begin{eqnarray}
M_i^{*}=A_i m-B_i - A_i g_{\sigma} \sigma,\quad i=p, n, t, h, d, \alpha,
\label{eq:Mj}
\end{eqnarray}%
where $g_{\sigma}$, $g_{\omega}$, and $g_{\rho}$ are coupling constants of $\sigma$, $\omega$, and $\rho$ mesons with nucleons, respectively; $A_i$ is mass number; $B_i$ is the in-medium cluster binding energy and $m$ is nucleon mass in vacuum which is taken to be $m=939$ MeV. It should be noted that here neutrons and protons are assumed to have the same mass in vacuum, but for astrophysical applications of nuclear matter EOS, experimental masses of neutrons ($m_{n}$) and protons ($m_{p}$) should be used for accuracy and this gives a linear term in the isospin dependence of nucleon mass. Nucleons form an isospin doublet with $\tau_3 \Psi_n = -\Psi_n$ and $\tau_3 \Psi_p = \Psi_p$. Similarly, for the triton and helium-3 one has $\tau_3 \Psi_t = -\Psi_t$ and $\tau_3 \Psi_h = \Psi_h$, respectively.

The meson Lagrangian densities are given by $\mathcal{L}_{\rm{meson}} = \mathcal{L}_{\sigma} + \mathcal{L}_{\omega}  + \mathcal{L}_{\rho} + \mathcal{L}_{\omega\rho}$ with
\begin{eqnarray}
&&\mathcal{L}_{\sigma}=\frac{1}{2}\partial_{\mu}\sigma\partial^{\mu}\sigma
-\frac{1}{2}m^2_{\sigma}\sigma^2-\frac{1}{3}g_2\sigma^3-\frac{1}{4}g_3\sigma^4, \\
&&\mathcal{L}_{\omega}=-\frac{1}{4}W_{\mu\nu}W^{\mu\nu}
+\frac{1}{2}m^2_{\omega}\omega_{\mu}\omega^{\mu}+\frac{1}{4}c_3\left(\omega_{\mu}\omega^{\mu}\right)^2,\\
&&\mathcal{L}_{\rho}=-\frac{1}{4}\overrightarrow{R}_{\mu\nu}\cdot\overrightarrow{R}^{\mu\nu}
+\frac{1}{2}m^2_{\rho}\overrightarrow{\rho}_{\mu}\cdot\overrightarrow{\rho}^{\mu},\\
&&\mathcal{L}_{\omega\rho}=\Lambda_v\left(g^2_{\omega}\omega_{\mu}\omega^{\mu}\right)
\left(g^2_{\rho}\overrightarrow{\rho}_{\mu}\cdot\overrightarrow{\rho}^{\mu}\right).
\label{eq:Lmeson}
\end{eqnarray}%
where $W^{\mu\nu}$ and $\overrightarrow{R}^{\mu\nu}$ are the antisymmetric field tensors for $\omega^{\mu}$ and $\overrightarrow{\rho}^{\mu}$, respectively. In the RMF approach, meson fields are treated as classical fields and the field operators are replaced by their expectation values.

In order to explore how the clustering effects depend on the NL-RMF interactions and the symmetry energy of nucleonic matter, we select four parameter sets of the NL-RMF model for nucleon degree of freedom, namely, NL3~\cite{Lal97}, FSU~\cite{Tod05}, FSUGold5~\cite{Pie06} and FSU-II~\cite{Cai12}. The parameter values of the four NL-RMF interactions are listed in Table~\ref{tab:1} for completeness. The FSUGold5 parameter set is obtained based on FSU by adjusting $g_{\rho}$ and $\Lambda_v$ as prescribed in Ref.~\cite{Pie06}, namely, for $\Lambda_v=0.05$ one readjusts only the $g_{\rho}$ to keep the symmetry energy $E_{\rm sym}(n_c)$ at $n_c = 0.1$ fm$^{-3}$ fixed. The parameters of FSU-II are taken from Ref.~\cite{Cai12} and they are obtained similarly as FSUGold5. One can then check the symmetry energy effects by adopting FSU, FSU-II and FSUGold5 with different values of $\Lambda_v$, which lead to different values of $E_{\rm sym}(n_0)$ as well as the density slope parameter $L=3n_0dE_{\rm{sym}}(n)/dn|_{n=n_0}$ of the symmetry energy. In particular, one has $E_{\rm sym}(n_0) = 30.6$ MeV and $L=45.8$ MeV for FSUGold5, $E_{\rm{sym}}(n_0) = 32.5$ MeV and $L=60.4$ MeV for FSU, and $E_{\rm{sym}}(n_0) = 35.5$ MeV and $L=87.4$ MeV for FSU-II. Compared with FSU, FSU-II and FSUGold5, the NL3 interaction additionally has significantly different isoscalar properties and thus can be used to test the interaction dependence of our results.

\begin{table}
\caption{Parameter sets of the NL-RMF Lagrangian used in this work.}
\label{tab:1}
\begin{tabular}{lc c c c}
\hline\hline
~ & NL3~\cite{Lal97} & FSU~\cite{Tod05} & FSUGold5~\cite{Pie06} & FSU-II~\cite{Cai12} \\
\hline
$m_{\sigma}$ (MeV)  &508.194  & 491.500  & 490.250 &491.500 \\
$m_{\omega}$ (MeV) &782.501  & 782.5    &782.5    & 782.5 \\
$m_{\rho}$ (MeV)   &763.0    & 763.0    &763.0    & 763.0 \\
$g_{\sigma}$        &-10.2170 & -10.5924 &-10.5924 &-10.5924  \\
$g_{\omega}$       &12.8680  & 14.3369  & 14.3369 & 14.3369 \\
$g_{\rho}$         &8.9480   & 11.7673 & 16.3739  & 9.6700 \\
$g_{2}$ (fm$^{-3}$) &10.4310  &4.2771   &4.2771  & 4.2771 \\
$g_{3}$           &-28.8850  & 49.8556  &49.8556 & 49.8556 \\
$c_{3}$           &0        & 422.4953 &422.4953& 422.4953 \\
$\Lambda_v$       &0        & 0.030   & 0.050    &0.010 \\
\hline\hline
\end{tabular}
\end{table}

The in-medium cluster binding energy $B_i=B_i^0+\Delta B_i$ is dependent on temperature $T$, total proton number density $n^{tot}_{p}$, and total neutron number density $n^{tot}_{n}$ of the system, where $B_i^0$ denotes the binding energy for cluster $i$ in vacuum. The total energy shift of a cluster in nuclear medium mainly includes the contribution from self-energy shift which is already contained in the cluster effective mass in the gNL-RMF model, the Coulomb shift which can be calculated from the Wigner-Seitz approximation and the Pauli shift which was evaluated in the perturbation theory with Jastrow and Gaussian approaches for light clusters $d$, $t$, $h$ and $\alpha$~\cite{Typ10}. Since the Coulomb shift is very small for the light clusters $d$, $t$, $h$ and $\alpha$ considered here and thus neglected in the present work. The energy shift $\Delta B_i$ is thus from the Pauli shift and it is assumed to have the following empirical quadratic form~\cite{Typ10}, i.e.,
\begin{eqnarray}
\Delta B_i(n^{tot}_{p}, n^{tot}_{n}, T)= -\tilde{n}_{i}\left[1+\frac{\tilde{n}_{i}}{2 \tilde{n}_{i}^{0}}\right]\delta B_{i}(T),
\label{eq:delbind}
\end{eqnarray}%
where $\tilde{n}_{i}$ stands for
\begin{eqnarray}
\tilde{n}_{i}= \frac{2}{A_i}\left[Z_i n^{tot}_{p} + N_i n^{tot}_{n}\right],
\label{eq:abb}
\end{eqnarray}%
in which $Z_i$ and $N_i$ are proton number and neutron number of the cluster $i$, respectively.
The density scale for cluster $i$ is given by
\begin{eqnarray}
\tilde{n}_{i}^{0}\left(T\right) = \frac{B_{i}^{0}}{\delta B_{i}\left(T\right)}.
\label{eq:denscale}
\end{eqnarray}%
The temperature dependence comes from $\delta B_{i}\left(T\right)$ defined by~\cite{Typ10}
\begin{eqnarray}
\label{eq:thashift}
&&\delta B_i\left(T\right)=\frac{a_{i, 1}}{\left(T+a_{i, 2}\right)^{3/2}}, \quad i=\alpha, t, h, \\
\label{eq:dshift}
&&\delta B_d\left(T\right) \\ \nonumber
&&=\frac{a_{i, 1}}{T^{3/2}}\left[\frac{1}{\sqrt{y_i}}-\sqrt{\pi}a_{i, 3}\exp\left(a_{i, 3}^2 y_i\right)\mathrm{erfc}\left(a_{i, 3}\sqrt{y_i}\right)\right],
\end{eqnarray}%
with $y_i=1+a_{i, 2}/T$. The values of  parameters $a_{i, 1}$, $a_{i, 2}$, and $a_{i, 3}$ are taken from Ref.~\cite{Typ10} and listed in Table~\ref{tab:2} for completeness.

\begin{table}
\caption{Parameters for the in-medium cluster binding energy shifts. The values are taken from Ref.~\cite{Typ10} and the values of $a_{\alpha, 1}$ and $a_{d, 1}$ in the parenthesis are the revised values in the present work to fit the experimental Mott densities~\cite{Hag12}.}
\label{tab:2}
\begin{tabular}{lc c c c c}
\hline\hline
Cluster $i$ & $a_{i , 1}$ & $a_{i, 2}$ & $a_{i, 3}$ & $B_i^0$ \\
~        & (MeV$^{5/2}$ fm$^{3}$)& (MeV) & (MeV) & (MeV) \\
\hline
$\alpha$  & 164371 (137330)  & 10.6701 &   -     & 28.29566 \\
$d$      & 38386.4 (76500) & 22.5204 & 0.2223  & 2.224566 \\
$t$      & 69516.2 & 7.49232 &   -     & 8.481798 \\
$h$      & 58442.5 & 6.07718 &   -     &7.718043 \\
\hline\hline
\end{tabular}
\end{table}

For homogeneous matter, the non-vanishing expectation values of meson fields are $\sigma=\langle\sigma\rangle$, $\omega=\langle\omega^0\rangle$, and $\rho=\langle\rho^3_0\rangle$. Since the cluster binding energy is density dependent, the equations of motion for the meson fields have the following form:
\begin{eqnarray}
\label{eq:sEoM}
m_{\sigma}^2\sigma &+& g_2 \sigma^2+g_3 \sigma^3 = \sum_{i=p, n, \alpha, d, t, h} g^i_{\sigma} n_i^s , \\
\label{eq:oEoM}
m_{\omega}^2\omega&+&c_3 \omega^3 + 2\Lambda_v g^2_{\omega} g^2_{\rho} \omega \rho^2 = \sum_{i=p, n, \alpha, d, t, h} g^i_{\omega} n_i \nonumber \\
&-& \sum_{i=\alpha, d, t, h}\frac{m_{\omega}^2}{2g_{\omega}}\left(\frac{\partial \Delta B_i}{\partial n^{\rm{ps}}_p}+\frac{\partial \Delta B_i}{\partial n^{\rm{ps}}_n}\right)n_i^s, \\
\label{eq:rEoM}
m_{\rho}^2\rho &+& 2\Lambda_v g^2_{\omega} g^2_{\rho} \omega^2 \rho = \sum_{i=p, n, t, h} g^i_{\rho} I_3^i n_i \nonumber \\
&-& \sum_{i=\alpha, d, t, h}\frac{m_{\rho}^2}{g_{\rho}}\left(\frac{\partial \Delta B_i}{\partial n^{\rm{ps}}_p}-\frac{\partial \Delta B_i}{\partial n^{\rm{ps}}_n}\right)n_i^s,
\end{eqnarray}%
where $n_i^s$ is the scalar density, $n_i$ is the vector density, isospin $I_3^i$ is equal to $1/2$ for $i=p, h$ and $-1/2$ for $i=n, t$, and the meson-cluster couplings are assumed to have the following forms,
\begin{eqnarray}
g^i_{\sigma}=A_i g_{\sigma},\quad
g^i_{\omega}=A_i g_{\omega},\quad
g^i_{\rho}=g_{\rho}.
\label{eq:coup}
\end{eqnarray}
In the above derivations, to avoid complications due to the total baryon density dependence of the cluster binding energies in the present theoretical framework, following the work of Typel \textit{et al}.~\cite{Typ10}, the dependence on the total baryon density in Eq.~(\ref{eq:delbind}) is replaced by a dependence on the pseudodensities which are defined by
\begin{eqnarray}
\label{eq:pseudo}
n^{\rm{ps}}_n=\frac{1}{2}\left[\rho_{\omega}-\rho_{\rho}\right], \quad
n^{\rm{ps}}_p =\frac{1}{2}\left[\rho_{\omega}+\rho_{\rho}\right],
\end{eqnarray}
with
\begin{eqnarray}
\label{eq:pseudoor}
\rho_{\omega}=\frac{m^2_{\omega}}{g_{\omega}}\sqrt{\omega^{\mu}\omega_{\mu}} ,\quad
\rho_{\rho}=\frac{2m^2_{\rho}}{g_{\rho}}\sqrt{\overrightarrow{\rho}^{\mu}\overrightarrow{\rho}_{\mu}}.
\end{eqnarray}%

The clusters are treated as point-like particles, and the vector and scalar densities of the fermions ($i=p, n, t, h$) are given, respectively, by
\begin{eqnarray}
\label{eq:fden}
n_i&=&g_i\int \frac{d^3k}{\left(2\pi\right)^3}\left[f_i^+(k)-f_i^-(k)\right],\\
\label{eq:fsden}
n_i^s&=&g_i\int \frac{d^3k}{\left(2\pi\right)^3}\frac{M_i^*}{\sqrt{k^2+M_{i}^{*2}}} \nonumber \\
&& \times \left[f_i^+(k)+f_i^-(k)\right],
\end{eqnarray}%
with degeneracy factor $g_i=2$ and the occupation probability given by the Fermi-Dirac distribution, i.e.,
\begin{eqnarray}
f_{i}^{\pm}=\frac{1}{1+\exp{\left[\left(\sqrt{k^{2}+M_{i}^{*2}}\mp\nu_{i}\right)/T\right]}}.
\label{eq:fermi}
\end{eqnarray}%
The densities of the bosons ($i=\alpha, d$) are obtained from
\begin{eqnarray}
\label{eq:bden}
n_i&=&g_i\int \frac{d^3k}{\left(2\pi\right)^3}\left[b_i^+(k)-b_i^-(k)\right],\\
\label{eq:bsden}
n_i^s&=&g_i\int \frac{d^3k}{\left(2\pi\right)^3}\frac{M_i^*}{\sqrt{k^2+M_{i}^{*2}}} \nonumber \\
&& \times \left[b_i^+(k)+b_i^-(k)\right],
\end{eqnarray}%
with degeneracy factor $g_{\alpha}=1$ and $g_d=3$, and the Bose-Einstein distribution gives the occupation probability in the following form:
\begin{eqnarray}
b_{i}^{\pm}=\frac{1}{-1+\exp{\left[\left(\sqrt{k^{2}+M_{i}^{*2}}\mp\nu_{i}\right)/T\right]}}.
\label{eq:boson}
\end{eqnarray}%
For a system including nucleons and light clusters in chemical equilibrium as we are considering in the present work, $\nu_i$ is the effective chemical potential which is defined as $\nu_i = \mu_i - g^i_{\omega}\omega - g^i_{\rho} I^i_3 \rho$, where the chemical potential of cluster $i$ is determined by
\begin{eqnarray}
\mu_i=N_i\mu_n+Z_i\mu_p.
\label{eq:mueq}
\end{eqnarray}%

The thermodynamic quantities of homogeneous matter are easily derived from the energy-momentum tensor. The energy density is given by
\begin{eqnarray}
\epsilon &=&\sum_{i=p, n, t, h}g_i\int\frac{d^3k}{\left(2\pi\right)^{3}}\sqrt{k^{2}+M^{*2}}\left( f_{i}^{+}+f_{i}^{-}\right) \nonumber \\
&&+ \sum_{i=d, \alpha}g_i\int\frac{d^3k}{\left(2\pi\right)^{3}}\sqrt{k^{2}+M^{*2}}\left( b_{i}^{+}+b_{i}^{-}\right) \nonumber \\
&&+ \frac{1}{2}m_{\sigma }^{2}\sigma ^{2}+ \frac{1}{3}g_{2}\sigma ^{3}+\frac{1}{4}g_{3}\sigma ^{4} \nonumber \\
&&-\frac{1}{2}m_{\omega }^{2}\omega ^{2}-\frac{1}{4}c_{3}\omega ^{4} -\frac{1}{2}m_{\rho }^{2}\rho ^{2} \nonumber \\
&&+ \sum_{i=p, n, t, h}\left(g^i_{\omega }\omega n_i + g^i_{\rho}\rho I^i_3 n_i\right)-\Lambda_v g^2_{\omega}g^2_{\rho} \omega^2 \rho^2,
\label{eq:energy}
\end{eqnarray}%
the pressure is obtained as
\begin{eqnarray}
p &=&\frac{1}{3}\sum_{i=p, n, t, h}g_i\int\frac{d^3k}{\left(2\pi\right)^{3}}
\frac{k^2}{\sqrt{k^{2}+M^{*2}}}\left(f_{i}^{+}+f_{i}^{-}\right) \nonumber\\
 &&+ \frac{1}{3} \sum_{i=d,\alpha}g_i\int\frac{d^3k}{\left(2\pi\right)^{3}}
\frac{k^2}{\sqrt{k^{2}+M^{*2}}}\left(b_{i}^{+}+b_{i}^{-}\right) \nonumber \\
&&- \frac{1}{2}m_{\sigma }^{2}\sigma ^{2}- \frac{1}{3}g_{2}\sigma ^{3}-\frac{1}{4}g_{3}\sigma ^{4} \nonumber \\
&&+\frac{1}{2}m_{\omega }^{2}\omega ^{2}+\frac{1}{4}c_{3}\omega ^{4} +\frac{1}{2}m_{\rho }^{2}\rho ^{2} \nonumber \\
&&+\Lambda_v g^2_{\omega}g^2_{\rho} \omega^2 \rho^2,
\label{eq:pressure}
\end{eqnarray}%
and the entropy density is expressed as
\begin{eqnarray}
s&=&-\sum_{i=p, n, t, h}g_i\int\frac{d^3k}{\left(2\pi\right)^{3}} \left[f_{i}^{+}\ln f_{i}^{+}\right. \nonumber \\
&&+\left( 1-f_{i}^{+}\right) \ln \left(1-f_{i}^{+}\right) + f_{i}^{-}\ln f_{i}^{-} \nonumber \\
&&+\left.\left( 1-f_{i}^{-}\right) \ln \left(1-f_{i}^{-}\right) \right] -\sum_{i=\alpha, d} g_i\int\frac{d^3k}{\left(2\pi\right)^{3}} \nonumber \\
&& \times \left[b_{i}^{+}\ln b_{i}^{+}-\left( 1+b_{i}^{+}\right) \ln \left(1+b_{i}^{+}\right)\right. \nonumber \\
&&+ \left.b_{i}^{-}\ln b_{i}^{-}-\left( 1+b_{i}^{-}\right) \ln \left(1+b_{i}^{-}\right) \right].
\label{eq:entropy}
\end{eqnarray}%
These thermodynamic quantities satisfy the Hugenholtz-van--Hove theorem, i.e.,
\begin{eqnarray}
\epsilon=Ts-p+\sum_{i=p, n, d, t, h, \alpha} \mu_i n_i.
\label{eq:HH}
\end{eqnarray}%
It is convenient to define the internal energy per baryon as $E_{\rm{int}}=\epsilon/n-m$ and free energy per baryon as
\begin{eqnarray}
 F=E_{\rm{int}}-T\frac{s}{n}.
\label{eq:free}
\end{eqnarray}%

The binding energy per baryon of isospin asymmetric nuclear matter may be expanded in powers of isospin asymmetry $\delta=\left(n^{tot}_n-n^{tot}_p\right)/\left(n^{tot}_n+n^{tot}_p\right)$ up to $4$th-order, and then one has 
\begin{eqnarray}
E_{\rm{int}}\left(n, \delta, T\right)&=&E_{\rm{int}}\left(n, 0, T\right)+E_{\rm{sym}}(n, T)\delta^2 \nonumber\\
&+& E_{{\rm{sym}}, 4}(n, T)\delta^4+\mathcal{O}(\delta^6),
\label{eq:esym}
\end{eqnarray}%
where the density- and temperature-dependent symmetry energy $E_{\rm{sym}}$ and the $4$th-order symmetry energy $E_{\rm{sym, 4}}$ are defined by
\begin{eqnarray}
\label{eq:esym2}
E_{\rm{sym}}(n, T)&=&\frac{1}{2}\left.\frac{\partial^2 E_{\rm{int}}}{\partial \delta^2}\right|_{\delta=0},\\
\label{eq:esym4}
E_{{\rm{sym}}, 4}(n, T)&=&\frac{1}{24}\left.\frac{\partial^4 E_{\rm{int}}}{\partial \delta^4}\right|_{\delta=0}.
\end{eqnarray}%
Similarly, one can expand the free energy per baryon $F$ and the entropy per baryon $S$ in the same manner, i.e.,
\begin{eqnarray}
\label{eq:fsym}
F\left(n, \delta, T\right)&=&F\left(n, 0, T\right)+F_{\rm{sym}}(n, T)\delta^2 \nonumber\\
&+& F_{{\rm{sym}}, 4}(n, T)\delta^4+\mathcal{O}(\delta^6), \\
\label{eq:ssym}
S\left(n, \delta, T\right)&=&S\left(n, 0, T\right)+S_{\rm{sym}}(n, T)\delta^2 \nonumber\\
&+& S_{{\rm{sym}}, 4}(n, T)\delta^4+\mathcal{O}(\delta^6).
\end{eqnarray}%
The symmetry free energy and the $4$th-order symmetry free energy are given, respectively, by
\begin{eqnarray}
\label{eq:fsym2}
F_{\rm{sym}}(n, T)&=&\frac{1}{2}\left.\frac{\partial^2 F}{\partial \delta^2}\right|_{\delta=0},\\
\label{eq:fsym4}
F_{{\rm{sym}}, 4}(n, T)&=&\frac{1}{24}\left.\frac{\partial^4 F}{\partial \delta^4}\right|_{\delta=0},
\end{eqnarray}%
while the symmetry entropy and the $4$th-order symmetry entropy are obtained, respectively, as
\begin{eqnarray}
\label{eq:ssym2}
S_{\rm{sym}}(n, T)&=&\frac{1}{2}\left.\frac{\partial^2 S}{\partial \delta^2}\right|_{\delta=0},\\
\label{eq:ssym4}
S_{{\rm{sym}}, 4}(n, T)&=&\frac{1}{24}\left.\frac{\partial^4 S}{\partial \delta^4}\right|_{\delta=0}.
\end{eqnarray}%

Furthermore, one can investigate the clustering effects on the symmetry energy, symmetry free energy and symmetry entropy by checking the parabolic laws from which the $E_{\rm{sym}}(n, T)$, $F_{\rm{sym}}(n, T)$ and $S_{\rm{sym}}(n, T)$ are, respectively, replaced by
\begin{eqnarray}
\label{eq:epara}
E_{\rm{para}}(n, T)&=&E_{\rm{int}}\left(n, \delta = 1, T\right)-E_{\rm{int}}\left(n,0, T\right),\\
\label{eq:fpara}
F_{\rm{para}}(n, T)&=&F\left(n, \delta = 1, T\right)-F\left(n, 0, T\right),\\
\label{eq:spara}
S_{\rm{para}}(n, T)&=&S\left(n, \delta = 1, T\right)-S\left(n, 0, T\right).
\end{eqnarray}%

\section{Results and Discussion}
\label{Result}

\subsection{Mott densities of light clusters}

Mott densities are the densities at which the in-medium binding energies of clusters defined by $B_i=B_i^0+\Delta B_i$ vanish. The experimental Mott densities for light clusters $d$, $t$, $h$ and $\alpha$ are obtained by analyzing the data in heavy-ion collisions~\cite{Hag12}.
Using the medium-dependent binding energy shift $\Delta B_i$ parameterized by Eqs.~(\ref{eq:delbind}), (\ref{eq:thashift}) and (\ref{eq:dshift}) with the total baryon densities replaced by the pseudodensities defined by Eq.~(\ref{eq:pseudo}), one can calculate the light cluster Mott densities at each temperature.

\begin{figure}[!hpbt]
\includegraphics[scale=0.4, clip]{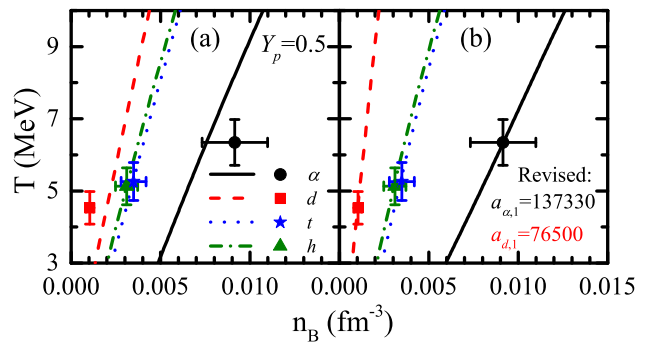}
\caption{The Mott density vs temperature in symmetric nuclear matter for light clusters $d$, $t$, $h$ and $\alpha$. The full symbols with error bars are experimental data from heavy-ion collisions by Hagel \textit{et al}.~\cite{Hag12} while the lines represent the predictions from the gNL-RMF model with original (panel (a))and revised (panel (b)) parameters for the in-medium light cluster binding energy shifts.}
\label{fig:Mott}
\end{figure}

\begin{figure}[!hpbt]
\includegraphics[scale=0.42, clip]{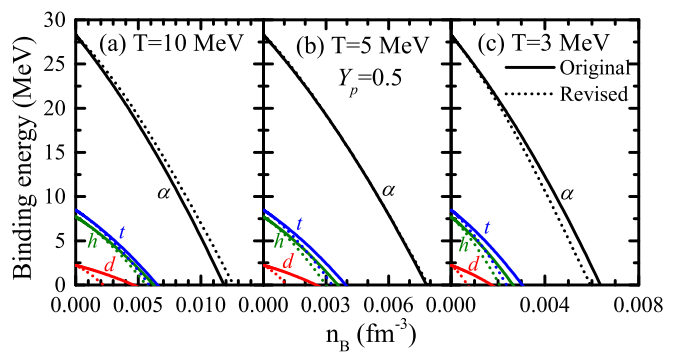}
\caption{Binding energies of light clusters $d$, $t$, $h$ and $\alpha$ as functions of the total baryon density $n_B$ at $T=10$ MeV (a), $5$ MeV (b) and $3$ (c) MeV. The results with the original~\cite{Typ10} and revised parameters (see Table~\ref{tab:2}) for the binding energy shifts are indicated by solid and dotted lines, respectively.}
\label{fig:Bind}
\end{figure}

Using the original parameters in Ref.~\cite{Typ10} as shown in Table~\ref{tab:2}, we display in Fig.~\ref{fig:Mott}(a) the calculated Mott densities for light clusters $d$, $t$, $h$ and $\alpha$. The corresponding experimental results from Ref.~\cite{Hag12} are also included for comparison. It is seen that the Mott densities increase with temperature. At a fixed temperature, the $\alpha$-particle has the largest Mott density and the next are the triton and helium-3, and the smallest is deuteron.
In addition, while the theoretical results agree well with the data for triton and helium-3, they significantly deviate from the data for $\alpha$-particle and deuteron. To fit the experimental data on the Mott densities of $d$ and $\alpha$, one can adjust the values of $a_{\alpha, 1}$ and $a_{d, 1}$ as well as $a_{\alpha, 2}$ and $a_{d, 2}$ in Eqs.~(\ref{eq:thashift}) and (\ref{eq:dshift}). In the present work, for simplicity, we only adjust the values of $a_{\alpha, 1}$ and $a_{d, 1}$ to fit the data. In particular, we note that changing the parameter $a_{\alpha, 1}$ from $164371$ to $137330$ and $a_{d, 1}$ from $38386.4$ to $76500$ can nicely reproduce experimental data on the Mott densities of $d$ and $\alpha$, as shown in Fig.~\ref{fig:Mott}(b).
These new revised values of $a_{\alpha, 1}$ and $a_{d, 1}$ for $d$ and $\alpha$ are included in Table~\ref{tab:2} as shown in the parenthesis.

Changing the parameters of in-medium binding energy shifts is a simple and direct approach to fit the experimental results. On the other hand, as Typel~\textit{et al} mentioned in Ref.~\cite{Typ10}, the parameters of binding energy shifts are determined by low-density perturbation theory from the unperturbed cluster wave functions. To be more consistent, it would be better to modify the QS model parameters to obtain new mass shifts, but this is certainly beyond the scope of the present work. In addition, the experimental results may depend on some model assumptions and thus could contain systematic errors. In the present work, for simplicity, following Ref.~\cite{Typ10} we assume that the in-medium light cluster binding energies follow the formulation in Eq.~(\ref{eq:delbind}) but depend on pseudodensities, which may lead to the necessity of readjusting the parameters in Eqs.~(\ref{eq:thashift}) and (\ref{eq:dshift}).

Shown in Fig.~\ref{fig:Bind} is the density dependence of the binding energy for light clusters $d$, $t$, $h$ and $\alpha$ at temperature $T=10$ MeV, $5$ MeV and $3$ MeV using the original and revised parameters as shown in Table~\ref{tab:2}.
It is seen that in general, at lower temperatures, the cluster binding energies drop faster as the baryon density increases. Furthermore, the binding energies of deuteron, triton, and helium-3 drop faster with density by using the revised parameters than by using the original parameters, and this is also the case for $\alpha$-particle at $T$=3 MeV. In Fig.~\ref{fig:Bind}(b), the two lines for $\alpha$-particle almost overlap. With the increment of temperature, the results of triton and helium-3 with the revised parameters get closer to the results with the original parameters and at the same time the $\alpha$-particle binding energy drops more and more slowly with the density. At $T=10$ MeV, the $\alpha$-particle binding energy with the revised parameters drops slower than that with the original parameters.
We note that the results of triton and helium-3 in both Fig.~\ref{fig:Mott} and Fig.~\ref{fig:Bind} are also slightly changed with the revised parameters $a_{\alpha, 1}$ and $a_{d, 1}$, and this is due to the fact that the baryon density in Eq.~(\ref{eq:delbind}) is replaced by the pseudodensity in Eq.~(\ref{eq:pseudo}).

Using the revised parameters and the original parameters shown in Table~\ref{tab:2} for the in-medium binding energy shifts of the light clusters allow us to explore the in-medium binding energy effects on the properties of low density nuclear matter with light clusters. In the following, we use the revised parameters for the in-medium binding energy shifts of the light clusters unless noted otherwise.

\subsection{Compositions of low density nuclear matter with light clusters}
\begin{figure}[!hpbt]
\includegraphics[scale=0.34, clip]{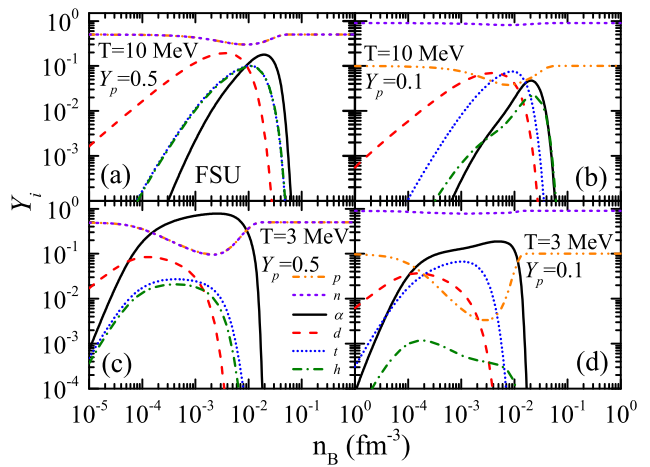}
\caption{The number fraction of nucleons and light clusters $d$, $t$, $h$ and $\alpha$ as a function of the total baryon density $n_B$ with FSU parameter set for $T=10$ MeV and $Y_p=0.5$ (a), $T=10$ MeV and $Y_p=0.1$ (b), $T=3$ MeV and $Y_p=0.5$ (c), and $T=3$ MeV and $Y_p=0.1$ (d).}
\label{fig:Frac}
\end{figure}

Shown in Fig.~\ref{fig:Frac} are the number fractions of nucleons and light clusters as functions of the total baryon density for isospin symmetric nuclear matter and neutron-rich nuclear matter with $Y_p=n^{tot}_{p}/n_B=0.1$ at temperature $T=3$ MeV and $10$ MeV with the FSU parameter set. It is seen that generally the deuteron dissolves first while the $\alpha$ dissolves last as baryon density increases for a fixed temperature and isospin asymmetry. This is understandable since the deuteron has smallest binding energy while the $\alpha$ has largest binding energy. For triton and helium-3, their dissolution densities are between those of deuteron and $\alpha$ and their fractions are almost identical for the isospin symmetric nuclear matter.

As the system becomes more neutron-rich, the fractions of $\alpha$, deuteron, and helium-3 become lower because of lacking in protons while the fraction of triton becomes higher than helium-3 as shown in Figs.~\ref{fig:Frac}(b) and~\ref{fig:Frac}(d).
As seen in Eq.~(\ref{eq:delbind}), the in-medium binding energies of triton and helium-3 are isospin-dependent with the binding energy of triton decreasing faster with increasing density while the binding energy of helium-3 drops more slowly for neutron-rich matter with $Y_p=0.1$. As a result, triton dissolves earlier while helium-3 dissolves later with increasing density in neutron-rich matter as shown in Figs.~\ref{fig:Frac}(b) and \ref{fig:Frac}(d). At lower temperature of $T=3$ MeV, among the light clusters, the $\alpha$-particle becomes the most dominant and dissolves last as shown in Figs.~\ref{fig:Frac}(c) and \ref{fig:Frac}(d). The $\alpha$ fraction is even larger than the nucleon fraction for symmetric nuclear matter around $n \sim 0.001$ fm$^{-3}$ at $T=3$ MeV, and thus the matter becomes the ``$\alpha$-matter''. At lower temperatures, nucleons prefer to form $\alpha$-particle since the $\alpha$ has the largest binding energy. As temperature increases, entropy becomes more and more important from the relation $F=E_{\mathrm{int}}-TS$. For isospin symmetric nuclear matter, instead of $\alpha$-particle, the nucleons become the most dominant at higher temperatures.
On the other hand, at lower temperatures, with increasing density, all clusters appear earlier at low density and dissolve earlier at high density since their binding energies decrease faster with increasing density at lower temperatures.

\begin{figure}[!hpbt]
\includegraphics[scale=0.42, clip]{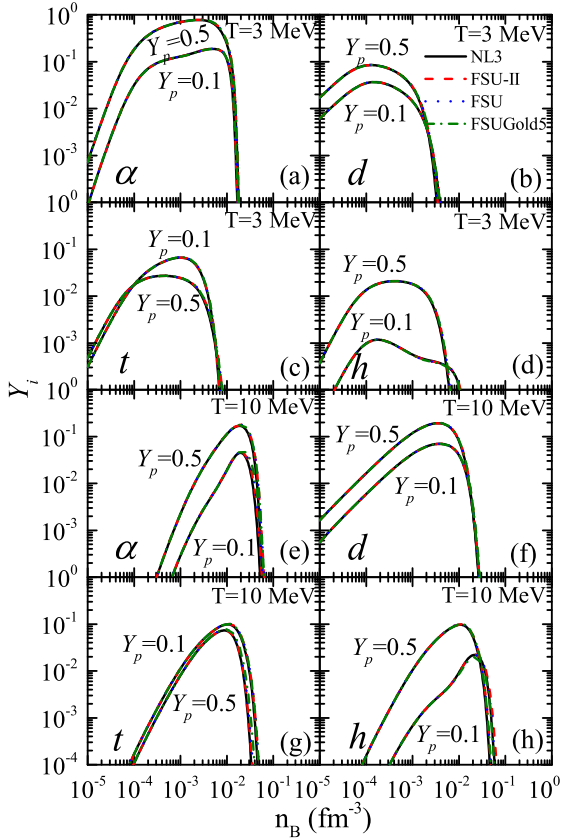}
\caption{The number fraction of light clusters $d$, $t$, $h$ and $\alpha$ as a function of the total baryon density $n_B$ in nuclear matter with $Y_p=0.5$ and $Y_p=0.1$ at $T=3$ MeV and $10$ MeV using the NL-RMF parameter sets NL3, FSU, FSUGold5 and FSU-II.}
\label{fig:FracRMF}
\end{figure}

The interaction-dependence of cluster fractions is checked in both symmetric nuclear matter and neutron-rich matter with $Y_p=0.1$ at $T=3$ MeV and $10$ MeV, and the results are shown in Fig.~\ref{fig:FracRMF}.
Four parameter sets, i.e., NL3, FSU, FSU-II and FSUGold5, are used for comparison. It is seen that the cluster fractions are almost identical in all cases, and thus they could be reasonably considered to be interaction-independent. This is mainly because the interactions among nucleons and clusters are relatively weak at low densities considered here.

\begin{figure}[!hpbt]
\includegraphics[scale=0.42, clip]{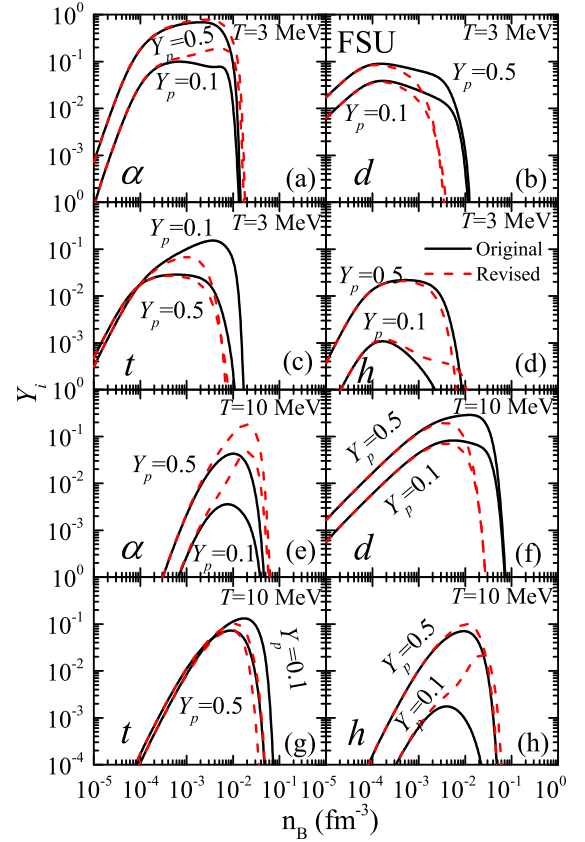}
\caption{The number fraction of light clusters $d$, $t$, $h$ and $\alpha$ as a function of the total baryon density $n_B$ in nuclear matter with $Y_p=0.5$ and $Y_p=0.1$ at $T=3$ MeV and $10$ MeV using the original~\cite{Typ10} and revised parameters (see Table~\ref{tab:2}) for the binding energy shifts.}
\label{fig:FracBin}
\end{figure}

Since the cluster fractions are essentially interaction-independent, it is thus interesting to see how the in-medium binding energy shifts influence the cluster fractions. We calculate the fractions of clusters with the original and revised parameters of binding energy shifts in various cases with the FSU parameter set and the results are shown in Fig.~\ref{fig:FracBin}.
It is seen that with increasing density, the $\alpha$-particle dissolves a little later and deuteron dissolves much earlier in all cases by using the revised parameters than using the original parameters. The difference between the results of triton and helium-3 is small in symmetric nuclear matter while become large in neutron-rich matter.

Therefore, our results indicate that the cluster fractions are essentially determined by the density- and temperature-dependence of the in-medium cluster binding energies. Furthermore, our calculations show that the light cluster fractions become important at low densities around $0.001$ fm$^{-3}$, especially at lower temperatures. In the density region of $n \gtrsim  0.02$ fm$^{-3}$, the fractions of light clusters in nuclear matter become insignificant and the nuclear matter is dominated by nucleons.

\subsection{Clustering effects on symmetry energy, symmetry free energy and symmetry entropy}

\begin{figure}[!hpbt]
\includegraphics[scale=0.44, clip]{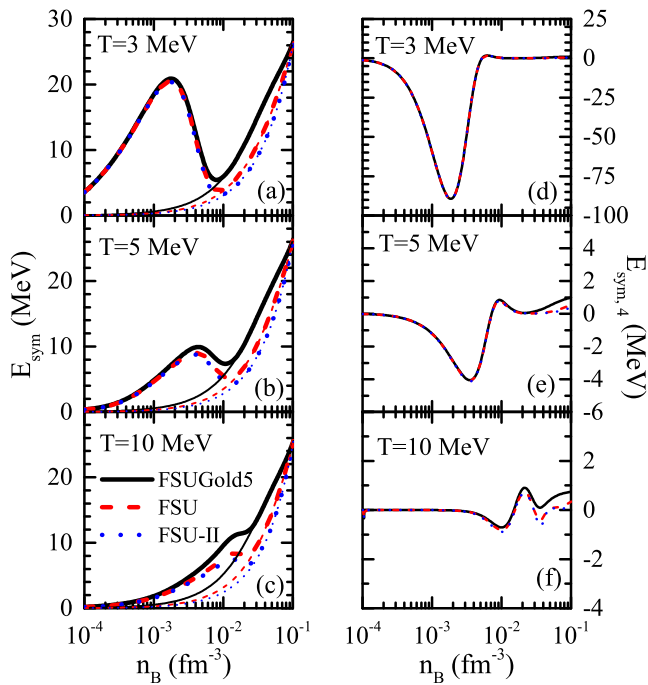}
\caption{Density dependence of $E_{\rm{sym}}$ (left panels) and $E_{\rm{sym, 4}}$ (right panels) at $T=3$ MeV, $5$ MeV and $10$ MeV in the gNL-RMF model with FSU, FSU-II and FSUGold5. For comparison, the results for $E_{\rm{sym}}$ in the NL-RMF model without considering clusters are also included (thin lines).}
\label{fig:EsymRMF}
\end{figure}

\begin{figure}[!hpbt]
\includegraphics[scale=0.44, clip]{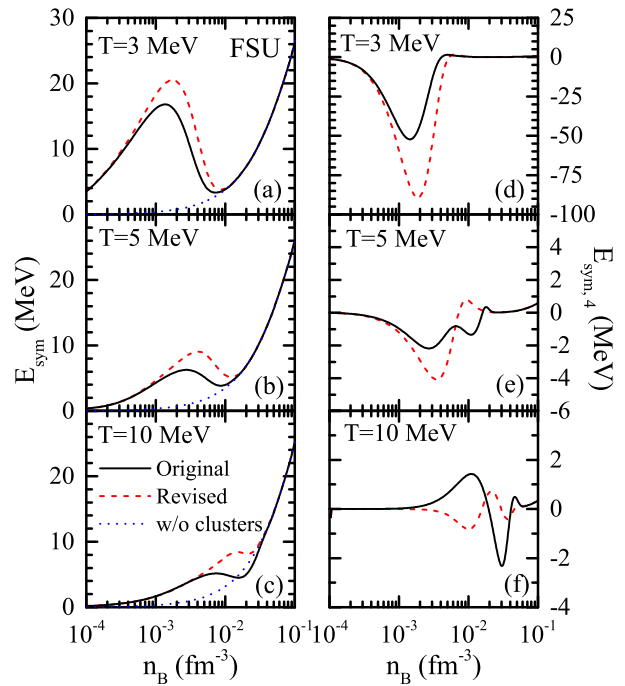}
\caption{Density dependence of $E_{\rm{sym}}$ (left panels) and $E_{\rm{sym, 4}}$ (right panels) at $T=3$ MeV, $5$ MeV and $10$ MeV in the gNL-RMF model using the original~\cite{Typ10} and revised parameters (see Table~\ref{tab:2}) for the binding energy shifts. For comparison, the results for $E_{\rm{sym}}$ in the NL-RMF model without considering clusters are also included (dotted lines).}
\label{fig:EsymBin}
\end{figure}

It is interesting to check the interaction-dependence of $E_{\rm{sym}}$ and $E_{\rm{sym, 4}}$ by using FSU, FSU-II and FSUGold5 which predicts essentially the same isoscalar properties but different density dependence of the symmetry energy by using different values of $\omega$-$\rho$ coupling $\Lambda_v$ and $\rho$ meson coupling $g_{\rho}$ as given in Table~\ref{tab:1}. Shown in Fig.~\ref{fig:EsymRMF} are $E_{\rm{sym}}$ and $E_{\rm{sym, 4}}$ as functions of the total baryon density at $T=3$ MeV, $T=5$ MeV and $T=10$ MeV with FSU, FSU-II and FSUGold5.
In all cases, it is seen that the $E_{\rm{sym}}$ are almost identical at very low densities, and then the differences begin to appear around $n_B \sim 0.003$ fm$^{-3}$ where the clustering effects become significant. At the higher density of $n_c=0.10$ fm$^{-3}$, the $E_{\rm{sym}}$ with different interactions meet each other, and this is because the symmetry energy values at $0.1$ fm$^{-3}$ are fixed for FSU, FSU-II and FSUGold5.
On the other hand, the $E_{\rm{sym, 4}}$ generally displays very small difference for difference interactions. It is interesting to see that the $E_{\rm{sym, 4}}$ becomes significantly negative at low densities and lower temperatures as shown in Fig~\ref{fig:EsymRMF}(d) and \ref{fig:EsymRMF}(e), and this breaks the empirical parabolic law of the symmetry energy which will be discussed in detail later.
Therefore, our results suggest that the interaction-dependence of $E_{\rm{sym}}$ and $E_{\rm{sym, 4}}$ at low densities is insignificant.

To investigate how the clustering effects influence the $E_{\rm{sym}}$, the corresponding results without considering clusters are also shown in Fig.~\ref{fig:EsymRMF} for comparison. It is clearly seen that the clustering effects significantly enhance the $E_{\rm{sym}}$ at low densities in all cases, and the enhancement becomes larger and larger with decreasing temperature.
On the other hand, the clustering effects disappear above about $0.01$ fm$^{-3}$ for $T=3$ MeV and above about $0.02$ fm$^{-3}$ for $T$=10 MeV for which the fraction of $\alpha$-particle begin to decrease as shown in Fig.~\ref{fig:FracRMF}. Similarly, the clustering effects also significantly enhance the $E_{\rm{sym,4}}$ at low densities in all cases (the $E_{\rm{sym,4}}$ without considering clusters is smaller than $0.5$ MeV~\cite{Cai12} in the density region considered in Fig.~\ref{fig:EsymRMF} and is not shown here), especially at lower temperatures.

Shown in Fig.~\ref{fig:EsymBin} are $E_{\rm{sym}}$ and $E_{\rm{sym, 4}}$ as functions of density using the FSU parameter set with the original and revised parameters for the in-medium binding energy shifts at $T=3$ MeV, $5$ MeV and $10$ MeV. One sees that the clustering effects with the revised parameters are stronger than those with the original parameters. Overall, the influences caused by different in-medium cluster binding energy shifts are more significant than that caused by different interactions as shown in Fig.~\ref{fig:EsymRMF}.
As can be seen by comparing the $E_{\rm{sym}}$ to that without clusters shown in Fig.~\ref{fig:EsymBin} by dotted lines, the density at which the clustering effects become negligible is essentially independent of the cluster binding energy shifts.

\begin{figure}[!hpbt]
\includegraphics[scale=0.34, clip]{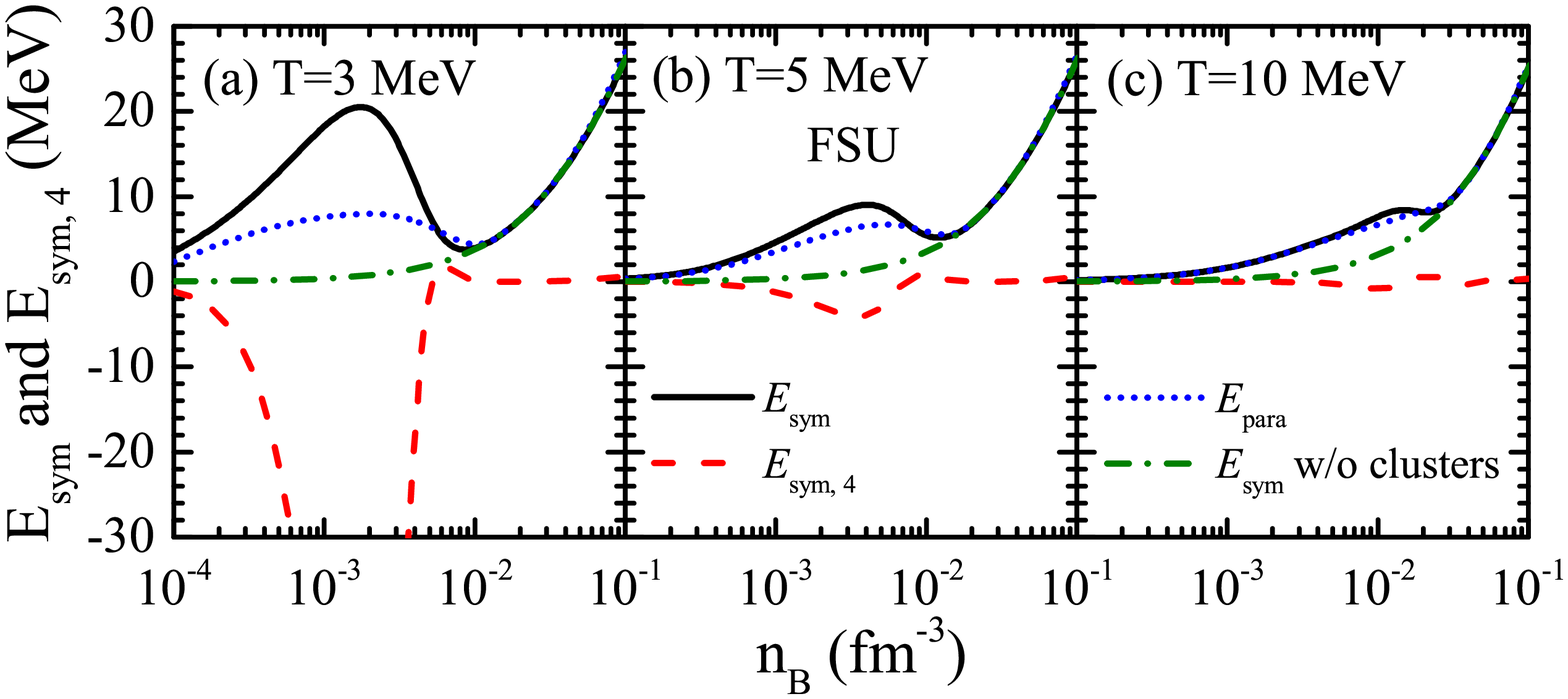}
\caption{Density dependence of $E_{\rm{sym}}$, $E_{\rm{sym, 4}}$ and $E_{\rm{para}}$ in the gNL-RMF model with FSU at $T=3$ MeV (a), $5$ MeV (b) and $10$ MeV (c). The results of $E_{\rm{sym}}$ in the NL-RMF model without considering clusters are also included for comparison.}
\label{fig:Esym}
\end{figure}

In order to explore more clearly about the clustering effects on the symmetry energy, we show in Fig.~\ref{fig:Esym} the $E_{\rm{para}}$ obtained by Eq.~(\ref{eq:epara}) as a function of the total baryon density at $T=3$ MeV, $5$ MeV and $10$ MeV by using FSU. For comparison, the results of $E_{\rm{sym}}$ and $E_{\rm{sym, 4}}$ obtained by Eqs.~(\ref{eq:esym2}) and (\ref{eq:esym4}) are also presented. One sees that the absolute values of $E_{\rm{sym, 4}}$ at low densities are relatively small at higher temperatures as shown in Fig.~\ref{fig:Esym}(c), and then get larger and larger with decreasing temperature. Focusing on Fig.~\ref{fig:Esym}(a) for $T=3$ MeV, one can see that there is a bulge for $E_{\rm{sym}}$ and valley for $E_{\rm{sym, 4}}$ at low densities where the light clusters are dominant. At low temperatures, the $\alpha$-particle is dominant and its large binding energy plays an important role in changing the symmetry energy. When the temperature increases, the nucleons become dominant and the entropy becomes important, therefore, the relatively small binding energies of clusters can hardly affect the symmetry energy. As a result, the clustering effects on the symmetry energy become weaker at higher temperatures.

From Fig.~\ref{fig:Esym}, one can see that the difference between the results of $E_{\rm{sym}}$ and $E_{\rm{para}}$ are much smaller than the absolute value of $E_{\rm{sym, 4}}$ at low densities and lower temperatures, which means that the expansion of internal energy per baryon in powers of isospin asymmetry is hard to get convergent, and thus the parabolic law for the isospin asymmetry expansion of nuclear matter EOS is invalid for nuclear matter including light clusters at low temperature, consistent with the conclusion in Ref.~\cite{Typ10}.
Furthermore, the results of $E_{\rm{sym}}$ without considering clusters are shown by dash-dotted lines in Fig.~\ref{fig:Esym} and the same conclusion is obtained as Fig~\ref{fig:EsymRMF}. Note that $E_{\rm{sym}}$ is much closer to $E_{\rm{para}}$ than the result without considering clusters at higher temperatures, and this means that the parabolic law could well approximate the $E_{\rm{sym}}$ at higher temperatures although the clustering effects slightly affect the symmetry energy.

\begin{figure}[!hpbt]
\includegraphics[scale=0.34, clip]{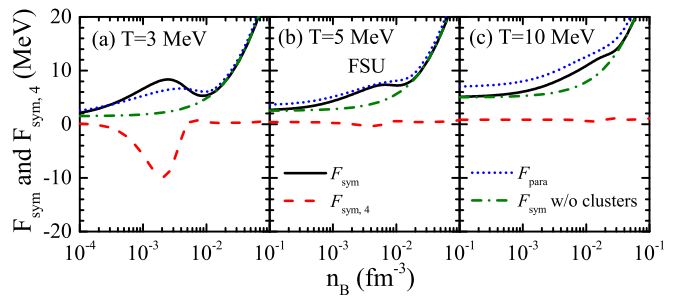}
\caption{Same as Fig.~\ref{fig:Esym}, but for the symmetry free energy.}
\label{fig:Fsym}
\end{figure}

\begin{figure}[!hpbt]
\includegraphics[scale=0.34, clip]{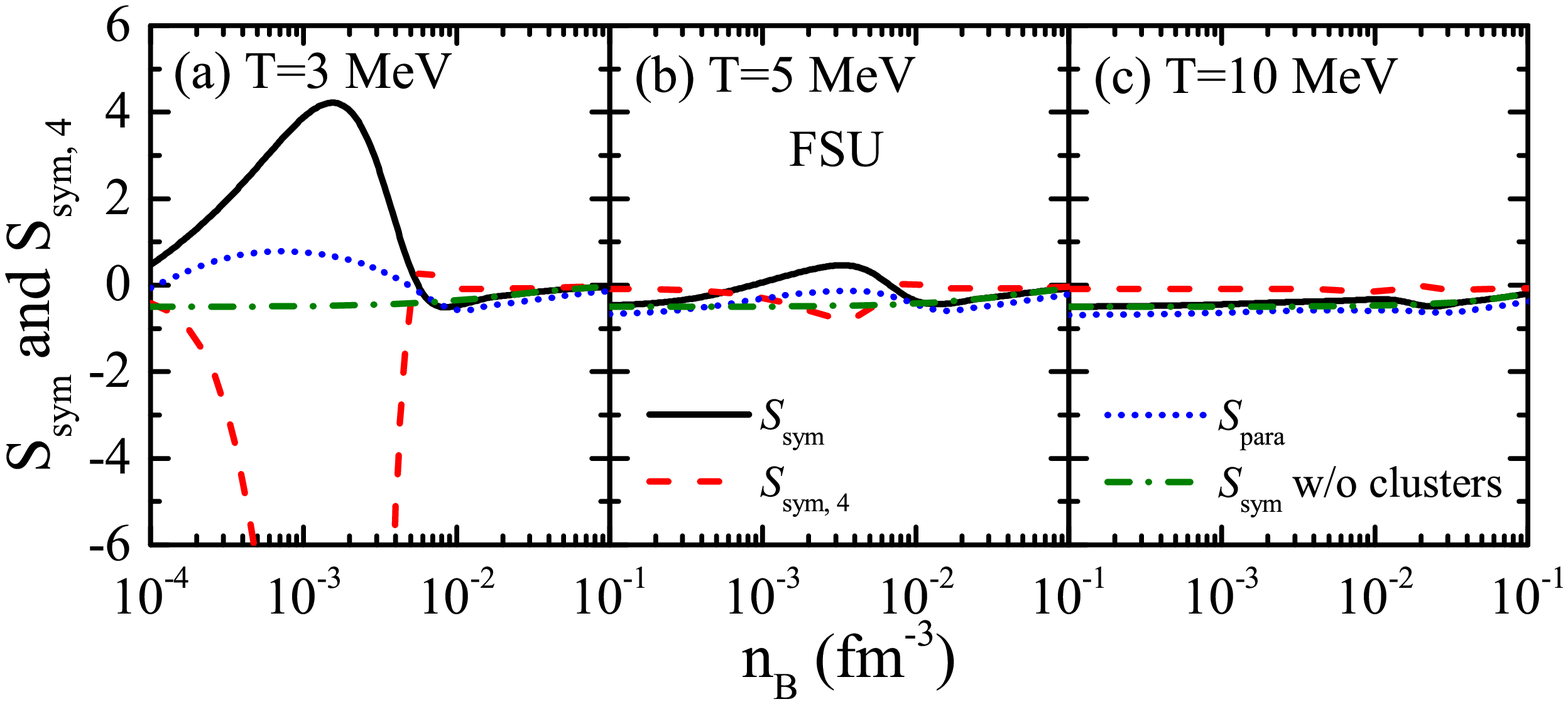}
\caption{Same as Fig.~\ref{fig:Esym}, but for the symmetry entropy.}
\label{fig:Ssym}
\end{figure}

Similarly to the symmetry energy, the results for the symmetry free energy and the symmetry entropy are shown in Fig.~\ref{fig:Fsym} and Fig.~\ref{fig:Ssym}, respectively. $F_{\rm{sym, 4}}$ and $S_{\rm{sym, 4}}$ are quite large at lower densities and lower temperatures, and then become smaller at higher temperatures. Focusing on Fig.~\ref{fig:Fsym}(a), one sees that the magnitudes of the bulge for $F_{\rm{sym}}$ and valley for $F_{\rm{sym, 4}}$ at low densities are smaller than those of $E_{\rm{sym}}$ and $E_{\rm{sym, 4}}$ as shown in Fig.~\ref{fig:Esym}(a), and this means that the parabolic law for free energy is broken not so strongly by the clustering effects compared with that for internal energy. As temperature increases, the clustering effects become weaker and weaker.

From Fig.~\ref{fig:Ssym}(a), one sees that the clustering effects on the symmetry entropy are significant at $T=3$ MeV, similar to the case of the symmetry energy as shown in Fig.~\ref{fig:Esym}(a). According to $F=E_{\mathrm{int}}-TS$, the relatively weaker clustering effects on the symmetry free energy observed in Fig.~\ref{fig:Fsym}(a) are thus mainly due to the significant cancellation between the symmetry entropy and the symmetry energy.
The results of $F_{\rm{sym}}$ and $S_{\rm{sym}}$ without considering clusters are also shown in Fig.~\ref{fig:Fsym} and Fig.~\ref{fig:Ssym}, and the conclusions obtained from these figures are very similar to that from $E_{\rm{sym}}$, namely, the clustering effects exist at lower densities and lower temperatures, and disappear at higher densities regardless of low or high temperatures.

\begin{figure}[!hpbt]
\includegraphics[scale=0.32, clip]{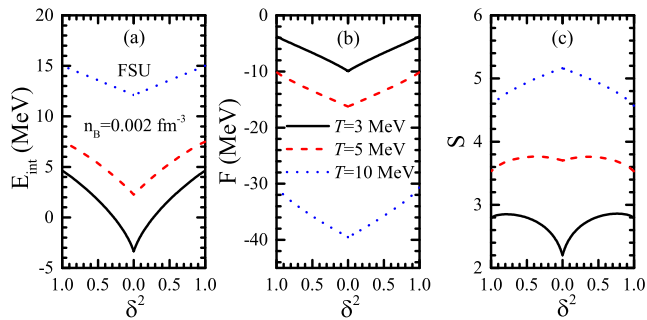}
\caption{$E_{\rm{int}}$ (a), $F$ (b), and $S$ (c) vs isospin asymmetry square ($\delta^2=\left(1-2Y_p\right)^2$) in the gNL-RMF model with the FSU parameter set at $n_B=0.002$ (fm$^{-3}$) and $T=3$ MeV, $5$ MeV and $10$ MeV.}
\label{fig:Delta}
\end{figure}

To intuitively illustrate how the clustering effects break the parabolic law for isospin asymmetry expansion of $E_{\rm{int}}$, $F$ and $S$, we present in Fig.~\ref{fig:Delta} the $E_{\rm{int}}$, $F$ and $S$ as functions of isospin asymmetry square $\delta^2=\left(1-2Y_p\right)^2$ at total baryon density $0.002$ fm$^{-3}$ (where the clustering effects are relatively strong as seen from previous figures) and $T=3$ MeV, $5$ MeV and $10$ MeV. It is seen that there are two symmetric branches as a function of $\delta^2$ with the left branch corresponding to the results from proton-rich matter calculations while the right branch from neutron-rich matter calculations. The nice symmetry of the two branches with respect to $\delta^2$ reflects the isospin symmetry breaking due to the small difference between the binding energies of triton and helium-3 is very small.

\begin{figure*}[!hpbt]
\includegraphics[scale=0.5, clip]{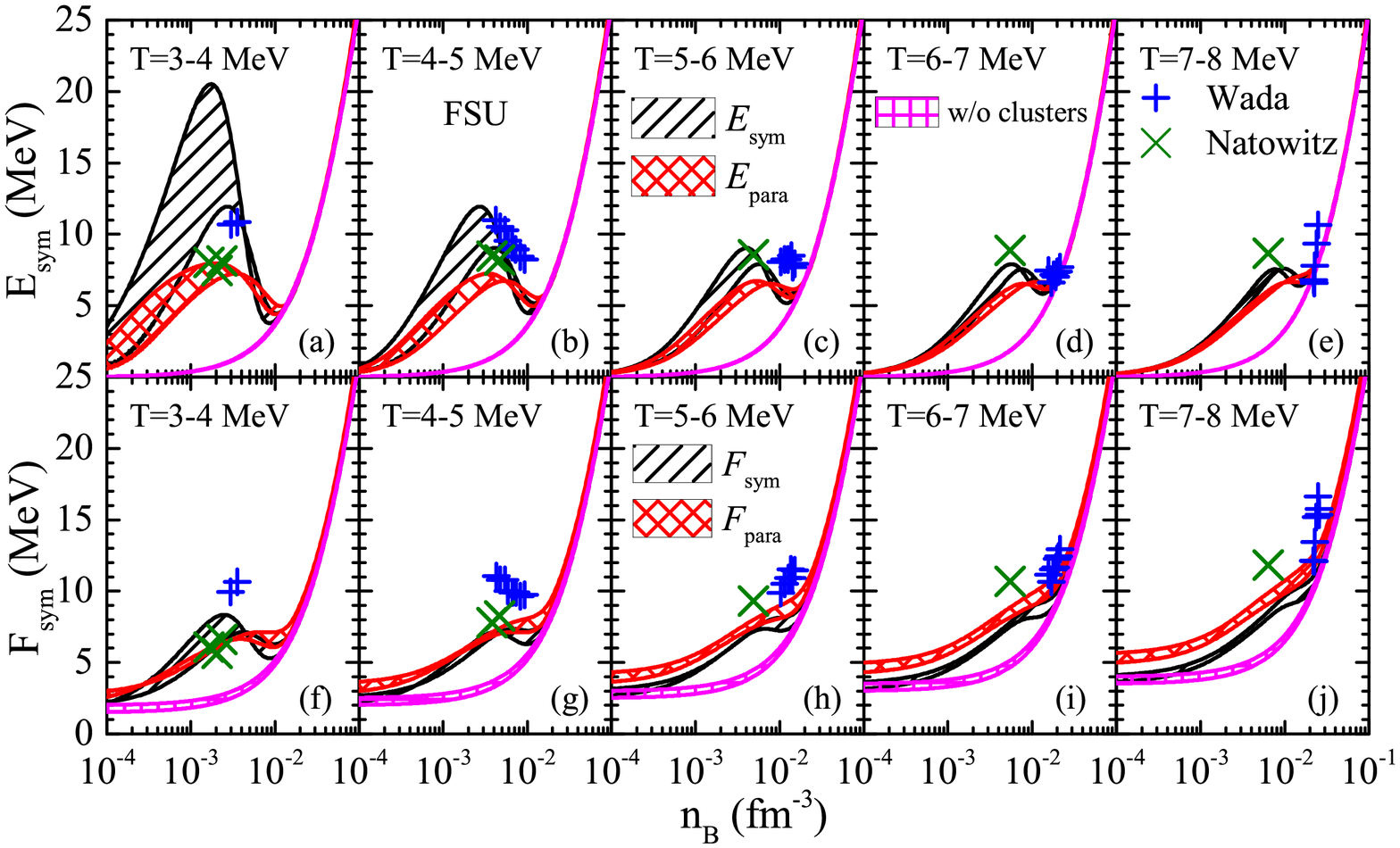}
\caption{The gNL-RMF model predictions for the density dependence of the symmetry energy (a--e) and symmetry free energy (f--j) with FSU at different temperature intervals. The experimental data from Ref.~\cite{Wad12} and~\cite{Nat10} as well as the results in the NL-RMF model without considering clusters are also included for comparison.}
\label{fig:Data}
\end{figure*}

Focusing on Fig.~\ref{fig:Delta}(a), one can see that the internal energy per baryon $E_{\rm{int}}$ increases as the temperature increases. At a fixed temperature, the symmetric nuclear matter has the minimum internal energy per baryon. Moreover, for each branch, one can see a nice linear relationship between the $E_{\rm{int}}$ and $\delta^2$ at higher temperature $T$=10 MeV while the linear relationship is broken at lower temperature $T=$3 MeV. These features clearly indicate that at total baryon density $0.002$ fm$^{-3}$, the parabolic approximation is broken for the isospin asymmetry expansion for the $E_{\rm{int}}$ at lower temperatures.
On the other hand, it is seen from Fig.~\ref{fig:Delta}(b) that the free energy per baryon $F$ decreases as the temperature increases, and for a fixed temperature it also reaches the minimum in symmetric nuclear matter. Compared with the results shown in Fig.~\ref{fig:Delta}(a) for the $E_{\rm{int}}$, the linear relationship between the $F$ and $\delta^2$ is broken not so much, and thus the parabolic law is approximately satisfied. This feature is consistent with the conclusion obtained from Fig.~\ref{fig:Fsym}.

As for the entropy, one can see from Fig.~\ref{fig:Delta}(c) that at $T=10$ MeV, the clustering effects are not important and the $S$ reaches its maximum value in symmetric nuclear matter and its minimum value in pure neutron (proton) nuclear matter as expected. It is interesting to see that at lower temperatures, e.g., $T=3$ MeV where the clustering effects become important, the $S$ reaches its minimum value in symmetric nuclear matter with a complicated dependence on the $\delta^2$. These features thus show that the clustering effects drastically influence the entropy per baryon in low density nuclear matter with light cluster at lower temperatures.

The above results demonstrate that the clustering effects play a significant role for the thermodynamic properties of low density nuclear matter, especially at lower temperatures. For low density nuclear matter at lower temperatures, the $4$th-order symmetry energy, the $4$th-order symmetry free energy and the $4$th-order symmetry entropy are found to be significant and the isospin asymmetry expansion for these nuclear matter properties is difficult to get convergent, indicating that the empirical parabolic law is invalid in this case.

\subsection{Comparison with data on symmetry energy and symmetry free energy}

Experimentally, the symmetry energy and the symmetry free energy at low densities and finite temperatures of $T \approx 3 \sim 8$ MeV have been extracted from analyzing the isoscaling behaviors of fragment production in heavy-ion collisions~\cite{Nat10,Wad12}. Shown in the upper (lower) panels of Fig.~\ref{fig:Data} are the predicted symmetry (free) energy $E_{\rm{sym}}$ ($F_{\rm{sym}}$) as a function of baryon density for five temperature intervals, namely, $T=3$-$4$ MeV, $T=4$-$5$ MeV, $T=5$-$6$ MeV, $T=6$-$7$ MeV and $T=7$-$8$ MeV, by using the FSU parameter set. For comparison, Fig.~\ref{fig:Data} also includes the corresponding theoretical results ($E_{\rm{para}}$  and $F_{\rm{para}}$) from the parabolic approximation (i.e., Eq.~(\ref{eq:epara}) and Eq.~(\ref{eq:fpara})) and the corresponding experimental data~\cite{Nat10,Wad12}. In addition, the corresponding results for $E_{\rm{sym}}$ and $F_{\rm{sym}}$ without considering clusters are also included for comparison.

Firstly, it is seen from Fig.~\ref{fig:Data} that when the baryon density is larger than about $0.02$ fm$^{-3}$, there are essentially no clustering effects on the symmetry (free) energy and the data can be nicely reproduced by the theoretical calculations (see, e.g., Fig.~\ref{fig:Data}(d), (e), (i) and (j) where the data around $0.02$ fm$^{-3}$ are available).
When the baryon density is below about $0.02$ fm$^{-3}$, the clustering effects become more and more important, especially for the symmetry energy at lower temperatures. These features are consistent with the results presented and discussed earlier.

From the upper panels of Fig.~\ref{fig:Data}, one can see the theoretical predictions on $E_{\rm{sym}}$ and $E_{\rm{para}}$ are quite similar at higher temperatures (i.e., $T\gtrsim 6$ MeV) but their difference becomes more and more significant as the temperature decreases. Moreover, it is seen that the experimental data on the symmetry energy can be reasonably described by the theoretical calculations by considering the light clusters, especially by the theoretical predictions of $E_{\rm{sym}}$. On the other hand, the theoretical predictions of $E_{\rm{sym}}$ without considering light clusters significantly underestimate the experimental data.

For the symmetry free energy, the theoretical predictions on $F_{\rm{sym}}$ and $F_{\rm{para}}$ are quite similar in all the cases considered here as shown in the lower panels of Fig.~\ref{fig:Data}. And the experimental data on the symmetry free energy can be reasonably described by the theoretical calculations by considering the light clusters, especially by the theoretical predictions of $F_{\rm{para}}$. Similarly to the case of the symmetry energy, the theoretical predictions of $F_{\rm{sym}}$ without considering light clusters significantly underestimate the experimental data.

Based on the above discussions, we conclude that our theoretical calculations with considering light clusters can reasonably reproduce the general behaviors of the symmetry energy and symmetry free energy extracted from experiments. These results suggest that the clustering effects play a very important role in describing the thermodynamic properties of low density nuclear matter with density below about $0.02$ fm$^{-3}$, especially at lower temperatures.

\section{Conclusion}
\label{Summary}

In the present work, using the generalized nonlinear relativistic mean-field (gNL-RMF) model, we have systematically explored the thermodynamic properties of homogeneous nuclear matter with light clusters at low densities and finite temperatures. In the gNL-RMF model, light clusters up to $\alpha $ ($1 \le A \le 4$) are included as explicit degrees of freedom and treated as point-like particles, the interactions among various particles are described by meson exchanges, and the in-medium effects on the cluster binding energies are considered by density- and temperature-dependent energy shifts with the parameters obtained by fitting the experimental Mott densities of the clusters extracted from heavy-ion collisions around Fermi energies.

Firstly, we have found that the composition of low density nuclear matter with light clusters is essentially determined by the density- and temperature-dependence of the in-medium cluster binding binding energies while the interactions among various particles play a minor role. In particular, our results indicate that the light cluster fractions become significant at low densities around $0.001$ fm$^{-3}$, especially at lower temperatures. On the other hand, in the density region of $n \gtrsim  0.02$ fm$^{-3}$, the fractions of light clusters in nuclear matter become insignificant and the nuclear matter is dominated overwhelmingly by nucleons.

Secondly, for nuclear matter at low densities ($n \sim 10^{-3}$ fm$^{-3}$) and low temperatures ($T \lesssim 3$ MeV), compared with the values of the conventional (second-order) symmetry energy, symmetry free energy and symmetry entropy, we have demonstrated that the $4$th-order symmetry energy, the $4$th-order symmetry free energy and the $4$th-order symmetry entropy are significant and the conventional isospin asymmetry expansion for these nuclear matter properties is hard to get convergent. These features imply that the empirical parabolic law is invalid and the concept of the conventional (second-order) symmetry energy becomes meaningless for the description of EOS of nuclear matter at these low densities and low temperatures. Therefore, to describe the EOS of nuclear matter with light clusters at low densities ($n \sim 10^{-3}$ fm$^{-3}$) and low temperatures ($T \lesssim 3$ MeV), full calculations (without isospin asymmetry expansion) are necessary.

Finally, we have compared the gNL-RMF model predictions of the symmetry energy and symmetry free energy at low densities and finite temperatures with the corresponding experimental data extracted from heavy-ion collisions. We have found that our theoretical calculations with considering light clusters can reasonably reproduce the general behaviors of the symmetry energy and symmetry free energy extracted from experiments. Moreover, our results indicate that the clustering effects can be negligible for nuclear matter with density above about $0.02$ fm$^{-3}$ but they play a very important role in describing the symmetry energy and symmetry free energy of low density nuclear matter with density below about $0.02$ fm$^{-3}$, especially at lower temperatures.

The present work has focused on the ideal infinite homogenous nuclear matter system with $6$ components, namely, neutrons, protons and light clusters including deuteron, triton, helium-3 and $\alpha $-particle, under thermal and chemical equilibrium without considering Coulomb interactions. For a more realistic system, one should additionally include heavier nuclei, and the nucleons, light clusters and heavy nuclei can interact with each other via meson-exchanges. In addition, the Coulomb interaction should be considered for charged particles. Furthermore, the system may include electrons under the conditions of electrically charge neutrality and chemical equilibrium. These studies are in progress and will be reported elsewhere.

\begin{acknowledgments}
This work was supported in part by
the Major State Basic Research Development Program (973 Program) in China under
Contract Nos. 2013CB834405 and 2015CB856904,
the National Natural Science Foundation of China under Grant Nos. 11625521, 11275125
and 11135011,
the Program for Professor of Special Appointment (Eastern Scholar) at Shanghai
Institutions of Higher Learning,
Key Laboratory for Particle Physics, Astrophysics and Cosmology, Ministry of
Education, China,
and the Science and Technology Commission of Shanghai Municipality (11DZ2260700).
\end{acknowledgments}

\end{document}